\documentclass[12pt]{article}
\newcommand{\beq}{\begin{eqnarray}}
\newcommand{\eeq}{\end{eqnarray}}
\newcommand{\drawsquare}[2]{\hbox{%
\rule{#2pt}{#1pt}\hskip-#2pt
\rule{#1pt}{#2pt}\hskip-#1pt
\rule[#1pt]{#1pt}{#2pt}}\rule[#1pt]{#2pt}{#2pt}\hskip-#2pt
\rule{#2pt}{#1pt}}

\newcommand{\PSbox}[3]{\mbox{\rule{0in}{#3}\includegraphics{#1}\hspace{#2}}}
\newcommand{\Yfund}{\raisebox{-.5pt}{\drawsquare{6.5}{0.4}}}
\newcommand{\Ysymm}{\raisebox{-.5pt}{\drawsquare{6.5}{0.4}}\hskip-0.4pt%
        \raisebox{-.5pt}{\drawsquare{6.5}{0.4}}}
\newcommand{\Yasymm}{\raisebox{-3.5pt}{\drawsquare{6.5}{0.4}}\hskip-6.9pt%
        \raisebox{3pt}{\drawsquare{6.5}{0.4}}}

\newcommand{\jref}[4]{{\it #1} {\bf #2}, #3 (#4)}

\newcommand{\NPB}[3]{\jref{Nucl.\ Phys.}{B#1}{#2}{#3}}

\newcommand{\PLB}[3]{\jref{Phys.\ Lett.}{#1B}{#2}{#3}}
\newcommand{\PR}[3]{\jref{Phys.\ Rep.}{#1}{#2}{#3}}
\newcommand{\PRD}[3]{\jref{Phys.\ Rev.}{D#1}{#2}{#3}}

\newcommand{\PRL}[3]{\jref{Phys.\ Rev.\ Lett.}{#1}{#2}{#3}}

\renewcommand{\theequation}{\thesection.\arabic{equation}}
\makeatletter
\def\vereq#1#2{\lower3pt\vbox{\baselineskip1.5pt \lineskip1.5pt
\ialign{$\m@th#1\hfill##\hfil$\crcr#2\crcr\sim\crcr}}}
\makeatother

\begin{document}

\begin{titlepage}
\begin{center}
\today     \hfill    LBNL-41659 \\
~{} \hfill UCB-PTH-98/19  \\
~{} \hfill hep-th/9804061\\

\vskip .3in

{\Large \bf Instantons in Partially Broken Gauge Groups\footnote{This 
work was supported in part by the U.S. 
Department of Energy under Contract DE-AC03-76SF00098, and in part by the 
National Science Foundation under grant PHY-95-14797.}}

\vskip 0.3in

{\bf Csaba Cs\'aki\footnote{Research fellow, Miller Institute for 
Basic Research in Science.} and Hitoshi Murayama\footnote{Supported in 
part by an Alfred P. Sloan Foundation Fellowship.}}

\vskip 0.15in

{\em Theoretical Physics Group\\
     Ernest Orlando Lawrence Berkeley National Laboratory\\
     University of California, Berkeley, California 94720}

\vskip 0.1in
{\rm and}
\vskip 0.1in

{\em Department of Physics\\
     University of California, Berkeley, California 94720}

\vskip 0.1in
{\tt  csaki@thwk5.lbl.gov, murayama@lbl.gov}

\end{center}

\vskip .25in

\begin{abstract}
We discuss the effects of instantons in partially broken gauge groups 
on the low-energy effective gauge theory.  Such effects arise when 
some of the instantons of the original gauge group $G$ are no longer 
contained in (or can not be gauge 
rotated into) the unbroken group $H$.  In 
cases of simple $G$ and $H$, a good indicator for the existence of 
such instantons is the ``index of embedding.''  However, in the 
general case one has to examine $\pi_3 (G/H)$ to decide whether there 
are any instantons in the broken part of the gauge group.  We give 
several examples of supersymmetric theories where such instantons 
exist and leave their effects on the low-energy effective theory.

\end{abstract}

\end{titlepage}

\newpage

\section{Introduction}
\setcounter{equation}{0}
\setcounter{footnote}{0}
Instanton effects~\cite{tHooft} play a major role in the low-energy dynamics 
of strongly 
interacting gauge theories.  Proper understanding of these 
effects~\cite{Shifman,Veneziano,Affleck,ADS} was very 
important for the recent advances in describing asymptotically free 
and finite
supersymmetric gauge theories~[6-27]. In particular, instantons
are used in several different ways: instanton effects in 
prepotentials~[15,23-26],
Affleck-Dine-Seiberg-type  (ADS) superpotentials~\cite{ADS} 
forcing the 
fields away from the origin of the moduli space. In most cases,
these ADS type superpotential terms appear only 
when the gauge group is completely broken.  However, Intriligator and Seiberg 
noted in a footnote in Ref.~\cite{SO} that in certain cases, when the
index of embedding\footnote{The index of embedding is defined in 
Section~\protect\ref{index}.} of the  
unbroken gauge group $H$ into the original gauge group $G$ is non-trivial, 
there can be instanton effects in the partially broken gauge group $G/H$ 
which has to be taken into account.

The aim of this paper is to clarify the issue of when instanton corrections in
partially broken groups become important. We explain in detail how the 
connection between the index of embedding and the instantons in the partially 
broken gauge groups noted in the  footnote in Ref.~\cite{SO} arises for simple
groups. For the more general case of semisimple groups, however, one has to
consider $\pi_3 (G/H)$ in order 
to decide whether such instanton corrections
can arise. We give 
several examples of theories with non-trivial embeddings both for simple groups
and product groups and study the 
effects of the instantons in the partially broken gauge group. 
 All of these examples are based on $N=1$ (or $N=2$) supersymmetric
gauge theories. The only reason for choosing supersymmetric examples
is that our understanding of the dynamics of these theories is much better
than for non-supersymmetric theories. We would like to stress, however, that
the general discussion of Section 2 is not restricted to supersymmetric
theories.

The paper is organized as follows: in Section 2 we discuss the issue
of instantons in partially broken gauge groups in general. 
For the case of simple groups we define the 
index of embedding and show that it is a good indicator
for the existence of $G/H$ instantons. 
Then we discuss the general case,
and show that $\pi_3(G/H)$ is the relevant quantity to signal the 
presence of $G/H$ instantons, and discuss how to calculate it
 using the exact homotopic sequence.
In Section 3 we show several examples of theories where $G/H$ instantons 
exist.  We discuss the effects of the $G/H$ instantons on the
low-energy dynamics in these theories. 
 Finally we conclude in Section 4. Appendix A contains the
proof of the connection between the index of embedding and $\pi_3 (G/H)$,
while in Appendix B we present an explicit example of a $Z_2$ instanton.

\newpage 

\section{General Considerations}
\setcounter{equation}{0}
\setcounter{footnote}{0}

\subsection{Instantons in Completely Broken Gauge
Groups\label{completely}}

Instantons are classical solutions of the four dimensional Euclidean 
equations of motion of the pure Yang-Mills theory,
\begin{equation}
        D_{\mu} F_{\mu\nu} = 0,
        \hspace{2cm} A_{\mu}(x) \rightarrow i U(x) \partial_{\mu} 
        U(x)^{\dagger} \; \; 
        \hbox{ for } \; \; |x| \rightarrow \infty .
\end{equation}
These solutions 
can be topologically characterized by a gauge-group element $U(x)$ at 
the space-time infinity $S^{3}$ which belongs to a non-trivial element of 
$\pi_3 (G)$,  
the third homotopy group of the gauge group $G$. The Higgs scalars (if there 
are any in the theory) are set to zero in the instanton solution. The
instantons are topologically stable and can be used 
for semi-classical expansion of the 
path integral.  The one-instanton solutions are characterized by 
their size $\rho$, their position, and their orientation in the gauge
group, with $2\mu_G$ parameters, where 
$\mu_G$ is the Dynkin index of the adjoint representation of the group
(for $SU(N)$ $\mu_G=2N$, thus there are $4 N$ parameters).  
In general, for an instanton with winding number $\nu$ there
are $2\nu \mu_G$ parameters needed to describe the solution. For example,
for $SU(2)$, the one instanton has 8 parameters, which consist of the
four coordinates describing the position of the instanton, one parameter
corresponding to the size of the instanton and three parameters which  
describe how the instanton is oriented inside the $SU(2)$ group. For 
$SU(3)$ there are 12 parameters, which are five for the position and size, 
and seven for rotating the instanton into $SU(3)$ (one of the 
eight $SU(3)$ generators leaves the instanton invariant).

Once the Higgs fields are turned on and the gauge group is broken, instantons 
are no longer exact solutions to the classical equations of motion, which
is in accordance with the expectation that in the Higgs phase any quantity
should behave as $e^{-v|x|}$ for large $x$, where $v$ is the Higgs expectation
value.  Consider, for example, a one-instanton configuration.  The
Euclidean action of the gauge field kinetic term is fixed as
$S_{YM}=8\pi^2/g^2$, while the action of the Higgs field,\footnote{It is
  convenient to choose the origin of the potential such that $V(v) = 
  0$.  This is automatically true in supersymmetric theories.}
\begin{equation}
        S_{H} = \int d^{4} x \left(  |D_{\mu} H|^{2} + V(H) \right) ,
\end{equation}
is minimized with $S_H \sim 8\pi^{2} v^2 \rho^2$ due to dimensional
analysis.
Therefore the total action $S_{YM}+S_{H}$ is minimized in the limit $\rho 
\rightarrow 0$ and hence there is no smooth classical field configuration 
in the one-instanton sector.
However, for $\rho < (g v)^{-1}$ ({\it i.e.}\/, $S_{H} < S_{YM}$), 
the approximate instanton solutions
\begin{eqnarray} &D_{\mu} F_{\mu\nu} = 0, \hspace*{1cm}
&A_{\mu}(x) \rightarrow i U(x) \partial_{\mu} 
U(x)^{\dagger}\; \; 
\hbox{ for } |x| \rightarrow \infty  \label{eq:pureYM}\\
&D_{\mu} D_{\mu} H - V'(H)= 0, \hspace*{1cm} 
&H(x) \rightarrow U(x) v \; \;
\hbox{ for }\; \;  |x| \rightarrow \infty \label{eq:Higgs}
\end{eqnarray}
(obtained by neglecting the current induced by the non-vanishing scalar field 
in the equation for $A_{\mu}$) can still be used for semi-classical
expansion of the path integral.  
The Higgs configuration 
$v$ is at the minimum of the potential $V$.
Solving these equations is identical to the following problem: for the 
fixed instanton background (\ref{eq:pureYM}), find the minimum of the
Euclidean action for the Higgs field $S_H$
with specified boundary condition.\footnote{The boundary condition in 
  Eq. (\ref{eq:Higgs}) is a consequence of the requirement $S_H <
  \infty$, which in turn requires $D_\mu H \rightarrow 0$, $V(H)
  \rightarrow 0$ for $|x| \rightarrow\infty$.}
Under this given instanton background with fixed size, scaling 
the Higgs field configuration to zero size does not make the action 
smaller and hence there must be a smooth non-trivial Higgs field 
configuration.  By expanding all fields with the instanton and 
Higgs field background, one can further include the one-loop effects 
of quantum fluctuations.  Then the integral over the instanton size 
should be performed.  The classical action grows for larger instanton size, 
which damps the integral at large 
$\rho$ as $e^{-8\pi^{2} \rho^{2} v^{2}}$, while the quantum effects prefer 
larger instanton size from the running of the gauge coupling in the 
instanton factor 
\begin{equation}
	e^{-8\pi^{2}/g^{2}(\rho)} 
	= e^{-8\pi^{2}/g^{2}(M)} (\rho M)^{b_{0}},
\end{equation}
in asymptotically free ($b_{0} >0$) theories.  Here, $M$ is the 
ultraviolet cutoff.  The balance between two  
effects results in a finite and well-defined result after the integral over 
the instanton size, with the main contribution from $\rho^{2} \simeq 
b_{0}/16 \pi^{2} v^{2}$.  
Therefore, non-vanishing expectation value of the Higgs field acts as an 
infrared cutoff in the size $\rho$ of the instanton. 
For larger instantons of $\rho > (g v)^{-1}$,
the approximate solutions cannot be trusted because $S_{H}$ becomes as 
large as $S_{YM}$, but this is not a 
problem because the larger instantons are suppressed due to the 
classical action $S_{H} \sim 8\pi^{2} \rho^{2} v^{2}$ and are justified {\it
  a posteriori}\/ as a self-consistent approximation for asymptotically 
free theories as long as $b_{0} g^{2}(\rho )/16 \pi^{2} < 1$. 

A more rigorous treatment of the instantons
in broken groups is to consider constrained instantons~\cite{Affleck},
that is to introduce a 
constraint into the Lagrangian which fixes the instanton size $\rho$. 
Then the modified equations will have exact solutions, which are called
constrained instantons.  The constraints are integrated over in the 
end to recover the original theory without the constraint.
For our purposes, however, it suffices that even in the
presence of non-vanishing Higgs fields the instantons remain 
approximate solutions which can be used for semi-classical expansion
of the path integral for $\rho g v < 1$. 
Such instantons in completely broken groups are responsible for the ADS type
superpotentials and many other dynamical effects in supersymmetric gauge 
theories.

Since the gauge group is completely broken, the effect of these instantons 
have to be taken into account when constructing the low-energy theory. The 
reason is that the effective theory is no longer a gauge theory, thus there 
are no instantons contained in the low-energy theory that could reproduce the 
effect of the original instantons. Therefore the effects of the instantons in 
the broken group $G$ have to be taken into account; for example the ADS
superpotential has to be added to the theory. 
Similarly, all the effects of the $G$ instantons have to be added 
to the low-energy effective theory when $G$ is not completely broken, but
the unbroken subgroup does not contain instantons anymore. This is for example
the case in $N=2$ theories, where the adjoint VEV breaks $G$ to $U(1)^r$, 
where $r$ is the rank of $G$. Since there are no $U(1)$ instantons,
the effects of the $G$ instantons have to be added to the low-energy
$U(1)^r$ prepotential~\cite{N=2Seiberg}.

\subsection{Instantons in Partially Broken Gauge Groups and the Index of
Embedding\label{index}}

Let us now consider the situation when the gauge group $G$ is only partially 
broken by a scalar VEV to a non-abelian subgroup $H$. In this case, both $G$ 
and $H$ contain instantons, and the question we want to answer is whether
any instanton corrections have to be added to the low-energy theory 
based on the gauge group $H$. The answer depends on whether or not
all effects of the original $G$ instantons can be reproduced by the
effects of the instantons in the unbroken group $H$. 
If all $G$ instantons are contained in $H$ (or at least can be 
gauge rotated into
$H$) then all 
information about instantons is still encoded in the effective $H$ theory
and no instanton corrections need to be taken into account. However, if
some of the $G$ instantons are not contained in $H$ (but instead in the
broken part $G/H$) then the effects of these ``$G/H$ instantons'' have to be 
added  to the effective $H$ theory. 

To understand when these effects can occur, let us consider 
the fermionic zero modes of a given representation in a one-instanton 
background when the group $G$ is simple. The number of zero modes coincides 
with the Dynkin index $\mu$  
of a given representation due to the Atiyah--Singer index 
theorem. The Dynkin index can be defined by
\beq
{\rm Tr}_{R}\, T_aT_b =\mu_{R} \delta_{ab},
\eeq
where the $T_a$'s are the generators in the given representation 
$R$ of the group $G$. 
This index $\mu_{R}$ is the number of fermionic zero modes in the one 
instanton background due to the index theorem,
once the generators for the fundamental representation 
have been properly normalized. (For classical groups this corresponds to
normalizing the generators such that the Dynkin indices of the fundamental 
representations of $SU(N)$ and 
$Sp(2N)$ are one and those of the vector representations of $SO(N) (N>3)$ are 
two). Now let us define the {\it index of embedding}, $\alpha$. 
Consider a simple 
group $G$ and one of its simple subgroups $H$. A representation
$R$ of the group $G$ has a decomposition under the $H$ subgroup  
\beq 
R\to R_1+R_2+\ldots + R_k.
\eeq
The index of embedding $\alpha$ is then given by
\beq
\alpha = \frac{\sum_{i=1}^k \mu_{R_i}}{\mu_R}.
\eeq
This index is an integer independent of the choice of $R$
and for most embeddings equals one. 

It is easy to see that this index is relevant to decide whether
there are any instanton effects in $G/H$ which one needs to take into
account. If the index is one, a given representation has
equal number of zero modes both in $G$ and in $H$. This suggests that 
there is a one-to-one correspondence between the instantons of $G$ and the
instantons of $H$, and no additional instanton effects besides the 
ordinary instanton effects in $H$ need to be taken into account.

However, if the index is 
bigger than one, a given representation has $\alpha$ times as many
zero modes in the one instanton background of $H$ than in the one instanton 
background of $G$. Therefore, the 't Hooft operator from the one-instanton 
of $H$ is, roughly speaking, the 't Hooft operator from the 
one-instanton of $G$ raised to the $\alpha^{{\rm th}}$ power.
This shows that the one-instanton of the $H$ theory
actually corresponds to an $\alpha$-instanton effect in $G$, and that the
$1,2,\ldots ,\alpha -1$ instantons of $G$ are missing from the $H$ theory.
The one instanton of $G$ would correspond to a ``$1/{\alpha}$'' instanton 
of $H$, which does not exist, and therefore any effects of these 
$1,2,\ldots ,\alpha -1$ instantons 
which do not decouple in the low-energy limit have to be added to the 
low-energy effective theory. Thus we find that, if the index of 
embedding is bigger than one, there are potential instanton contributions 
from $G/H$ which need to be added to the low-energy effective theory~\cite{SO}.

Another consequence of the non-trivial index is a modified matching 
condition of the gauge coupling constants.  
One has to match the gauge couplings 
of the high- and low-energy theories as 
\begin{equation}
\label{couplings}
	\frac{1}{g_G^2}=\frac{1}{\alpha g_H^2} + \mbox{threshold corrections}
\end{equation}
in the case that the index of embedding $\alpha$ is non-trivial, due to the 
fact that the normalization of generators changes.

The non-trivial matching of the gauge coupling constants results in a 
non-trivial scale matching relation.  In the case of supersymmetric theories,
there is no threshold correction in the $\overline{\mbox{DR}}$ 
scheme at the one-loop level~\cite{AKT}, and furthermore the running of the 
holomorphic gauge coupling constant 
is one-loop exact due to holomorphy~\cite{SV,Nima}.  In theories with 
non-vanishing $\beta$-function at the one-loop level, this statement 
is true even non-perturbatively \cite{Nima,Bogdan}.  
Then the matching between the scales can be written down exactly.  
The usual scale matching relation for the
breaking $G\to H$ (if $\alpha=1$) is given by 
\beq 
\frac{\Lambda^{b_G}}{v^{b_G}}=\frac{\Lambda^{b_H}}{v^{b_H}},
\eeq
where $v$ is the Higgs VEV of the breaking of
$G$ to $H$.  Note that the one-instanton effects in a given theory are
proportional to $\Lambda^{b}$, where $b$ is the coefficient of the
one-loop $\beta$-function and $\Lambda$ is the dynamical scale of the theory. 
Therefore, this matching relation can also be interpreted as an 
expression of the equivalence between the one-instanton of the original 
$G$ theory and the one-instanton of the $H$ theory.
However, if the index $\alpha$ is bigger than one, then the
one-instanton factor of the low-energy $H$ theory should not be matched
to the one instanton factor of $G$, but to the $\alpha$ instanton factor,
and thus the matching should be modified to 
\beq
\label{matching}
\left(\frac{\Lambda^{b_G}}{v^{b_G}} \right)^{\alpha}=
\frac{\Lambda^{b_H}}{v^{b_H}}.
\eeq 
This is indeed what follows from the matching of the gauge 
coupling constants (\ref{couplings}).  

As the first example for the index of embedding, consider breaking
$SU(N)$ to $SU(N-1)$ by giving an expectation value to a field transforming 
in the fundamental of $SU(N)$.
In this case the index of embedding is one. This can be seen by
considering the fundamental representation of $SU(N)$.  Its
decomposition under $SU(N-1)$ is given by $ \Yfund \to \Yfund +1$,
and since the Dynkin indices of the fundamental representations
of $SU(N)$ and of $SU(N-1)$ are both one, the index of embedding is one. 

However, if we consider the breaking $SU(N)\to SO(N)$,
the index will be non-trivial. This breaking can be achieved for example 
by giving an expectation value to a rank-two symmetric tensor of $SU(N)$. 
The fundamental representation of $SU(N)$, which has Dynkin index one,
will turn into the 
vector representation of $SO(N)$, which has Dynkin index two (for
$N>3$). Thus in this case 
the index of embedding is $\alpha =2$. Therefore, in this example, the 
one instanton of $SU(N)$ is missing from the $SO(N)$ theory, and the potential 
effects of this instanton have to be added to the low-energy theory.
For $N=3$ the index of the
embedding $SU(3)\to SO(3)$ is instead four, which can be seen by considering 
the decomposition of the fundamental of $SU(3)$, which has Dynkin index one.
The fundamental representation of $SU(3)$ will turn into the vector
representation of $SO(3)$, however since 
$SO(3)$ is locally isomorphic to $SU(2)$, the vector representation of
$SO(3)$ is nothing but the  
adjoint representation of $SU(2)$, which has Dynkin index four. Thus
$\alpha =4$ in this case. 
We will see an example of the effect of these $G/H$ instantons in 
Section~\ref{SU3}. 

A further example of a non-trivial embedding is $Sp(2N)\to SU(N)$ which 
has index two. This breaking can occur when the rank-two symmetric
tensor (the adjoint) 
of $Sp(2N)$ obtains an expectation value. One way to see that the index is two
is to note that the fundamental of $Sp(2N)$ decomposes as
$\Yfund \to \Yfund +\overline{\Yfund}$ under $SU(N)$. Examples of the effects 
of these instantons will be discussed in Sections~\ref{N=2Sp} and~\ref{Sp}.

We argued that for simple groups 
the index of embedding $\alpha$ is a good indicator of whether 
instantons in $G/H$ exist.
For semisimple groups however, the index of embedding is ambiguous.
Consider for example $SU(N)\times SU(N)$ broken to the diagonal 
$SU(N)$. The representation $(\Yfund ,\overline{\Yfund})$ becomes 
an adjoint of the diagonal $SU(N)$. The Dynkin index of $(\Yfund ,
\overline{\Yfund})$ is $N$, while the Dynkin index of the adjoint of $SU(N)$ is
$2N$, so one would conclude that the index of embedding is $2$. 
However, if one considers the representation $(\Yfund ,1)$ of $SU(N)\times
SU(N)$, one would conclude that the index of embedding is one. Thus
the naive definition of the index of embedding for semisimple groups is 
not well-defined. Instead of insisting on finding a good generalization
for the index of embedding for semisimple groups, we will go in a 
different direction, and examine the third homotopy group 
$\pi_3 (G/H)$. We will show in Section \ref{subsec:connection},
that in the case of 
simple groups, there is a simple connection between $\pi_3 (G/H)$ and
the index of embedding. However, for semisimple groups   $\pi_3 (G/H)$
is still well-defined, and will be the indicator for the existence of
$G/H$ instantons in the general case as shown in Section~\ref{sec:condition}. 

It is easy to
understand in terms of the instantons why one has to go beyond the index of
embedding for the case of semisimple groups in order to decide whether
there are any instantons in the broken part of the gauge group.
Consider the above example of  $SU(N)\times SU(N)$ broken to the diagonal 
$SU(N)$.
A one instanton effect in the diagonal $SU(N)$ group
corresponds to a one instanton effect in the $SU(N)\times SU(N)$ group as well,
but it is a particular combination of the one instanton in the 
first $SU(N)$ factor and the one instanton of the second $SU(N)$ factor
(the $(1,1)$ instanton).
However, the one-instanton of the first $SU(N)$ factor (the $(1,0)$ 
instanton) is not contained in the diagonal  $SU(N)$. Similarly, the
$(0,1)$ instanton is not contained in the diagonal $SU(N)$ either, but this
instanton is equivalent to $(-1,0)$, the anti-instanton in the first $SU(2)$
factor.
This can be seen because
$(1,0)=(0,-1)+(1,1)$, where $(0,-1)$ is the anti-instanton in the 
second $SU(2)$ factor, therefore $(0,-1)$ is equivalent to $(1,0)$, and
so $(0,1)$ is equivalent to $(-1,0)$.
 Thus even though the one-instanton effect of the 
diagonal $SU(N)$ corresponds to one-instanton effects in both
$SU(N)$ factors, there are still 
$G/H$ instantons whose effects have to be taken into account.

To summarize this section, we have seen that for simple groups the 
index of embedding is a good indicator of whether $G/H$ instantons exist.
However, for product groups one has to rely on a different
analysis. In order to establish the connection between the index of 
embedding and
the existence of $G/H$ instantons for simple groups and to examine the
cases of non-simple groups we will need to examine the possible
topologies of the field configurations which could give rise to $G/H$
instantons. This is the subject of the next section.

\subsection{Instantons in Partially Broken Gauge Groups and $\pi_{3} (G/H)$
\label{sec:condition}}

We have seen in the previous section that for certain non-trivial embeddings 
of $H$ into $G$ the mapping between the instantons of $G$ and $H$ 
may be non-trivial which could affect the low-energy theory. In this
section we will consider the topology of these embeddings in order to decide 
whether $G/H$ instantons exist. The usual instantons are topologically 
stable, because there is a non-trivial mapping from the sphere at the 
infinity of space-time to the gauge group. This mapping $S^3 \to G$
is  characterized by the third homotopy group $\pi_3 (G)$. If we are 
interested whether any of these instantons are contained in $G/H$ instead 
of $H$ we have to ask whether $\pi_3 (G/H)$ is trivial.
This is because we expect that, just like in the case of the instantons
in completely broken gauge groups discussed in Section~\ref{completely},
the approximate instanton solution obtained by ignoring the Higgs 
field will still contribute to the path integral, and generate 
't Hooft operators. 

We now show that the non-trivial instantons which appear in 
the $G\rightarrow H$ breaking can be classified by $\pi_{3} (G/H)$.  
We again study the approximate field equations,
\begin{eqnarray}
        &D_{\mu} F_{\mu\nu} = 0,
        \hspace{1cm} & A_{\mu}(x) \rightarrow i U(x) \partial_{\mu} 
        U(x)^{\dagger} \; \; 
        \hbox{ for }\; \;  |x| \rightarrow \infty, \label{eq:pureYM2} \\
        & D_{\mu} D_{\mu} H - V'(H)= 0, \hspace*{1cm}
 &H(x) \rightarrow U(x) v \; \;
        \hbox{ for } \; \; |x| \rightarrow \infty . \label{eq:Higgs2}
\end{eqnarray}
Solving these equations is identical to the following problem: for the 
fixed instanton background (\ref{eq:pureYM2}), find the minimum of the
Euclidean action for the Higgs field 
\begin{equation}
        S_{H} = \int d^{4} x \left(  |D_{\mu} H|^{2} + V(H) \right) ,
\end{equation}
with specified boundary condition.  We again choose the 
origin of the potential such that $V(v) = 0$ at the minimum.

The gauge-group element $U(x) \in \hbox{Map}(S^{3} \rightarrow G)$ 
belongs to a non-trivial homotopy class in $\pi_{3} (G)$.  If $\pi_{3}
(G/H)$ is trivial, however,  
$U(x)$ can be continuously deformed to a gauge-group element 
$U_{H}(x)$ which lives purely in $H$, {\it i.e.}\/ $U(x)_{H} v = v$. 
By continuously deforming the Higgs 
field configuration from $U(x) v$ to $v$ at the space-time infinity, 
the boundary condition of the Higgs field is topologically trivial.
This continuous 
deformation can be done at negligible cost in the size of the action 
by making the deformation arbitrarily slow at infinity
\cite{Coleman-instantons}. Since,   
for one-instanton solutions, the gauge field configurations can also be
gauge-rotated to be contained in the $H$ part only, the Higgs 
field does not interact with the instanton solution any more and hence 
the configuration can be extended all the way to the center of the 
instanton, with vanishing action $S_{H} = 0$.  Then the 
field configuration is nothing but the instanton in the unbroken 
group $H$, where the Higgs field responsible for $G\rightarrow H$ 
breaking is frozen at the minimum of the potential.  The effects of
such a configuration should certainly not be explicitly 
included in the action of the low-energy effective $H$ theory because 
they are yet-to-be included in the dynamics of the low-energy $H$ theory.

On the other hand, if $\pi_{3} (G/H)$ is non-trivial, the Higgs field 
configuration $U(x) v$ at the space-time infinity cannot be ``unwound'' 
to a trivial configuration $v$.  Therefore, there must be a field 
configuration which minimizes the action $S_{H}$ in a given 
non-trivial class of $\pi_{3} (G/H)$ with $S_H \sim 8\pi^{2}v^2 \rho^2$.  In
this case, the field  
configuration involves the Higgs field in an essential manner, and 
such a configuration does not belong to the low-energy $H$ theory.  
The effect of this type of field configurations has to be included 
when writing down the effective action of the low-energy $H$ theory.
An explicit example of an instanton in a partially broken gauge group 
is presented in Appendix B.

The above argument strongly resembles that for a `t Hooft--Polyakov 
monopoles in three spatial dimensions (see, {\it e.g.}\/,
\cite{Coleman-monopoles,Preskill-monopoles}).  One 
important difference,  
however, is that it is possible to further decrease the size of the 
action by scaling {\it both}\/  the Higgs field and gauge field 
configurations to zero size.  Note that 
we are keeping the instanton 
background fixed in the argument; under this given background, scaling 
the Higgs field configuration to zero size does not make the action 
smaller and hence there must be a smooth non-trivial Higgs field 
configuration.  By expanding all fields with the instanton and 
Higgs field background, the classical action grows as $8\pi^{2} \rho^{2} 
v^{2}$ for larger instanton size, while the quantum effects prefer 
larger instanton size from the running of the gauge coupling in the 
instanton factor $e^{-8\pi^{2}/g^{2}(\rho)}$ if the breaking of the
gauge group makes the coupling less asymptotically free.  The balance
between the two effects results in finite well-defined result after the
integral over the instanton size.  This is a trivial extension of
the argument as in the completely broken gauge theories.

\subsection{The Index of Embedding and $\pi_3(G/H)$\label{sec:pi3}}
\label{subsec:connection}

We have seen that the non-trivial Higgs field configuration 
under an instanton background can be classified according to  
$\pi_{3} (G/H)$.  We have also seen earlier that the index of embedding 
has something to do with the presence of non-trivial $G/H$ instanton 
in a heuristic manner by using the index theorem and the number of fermion 
zero modes.  In this subsection, we would like to see the connection 
between the two arguments, and see that the argument based on $\pi_{3}
(G/H)$ reduces to that based on the index of embedding if both the
groups $G$ and $H$ are simple.

Let us consider first the case when both $G$ and $H$ are simple groups.
We have seen in the previous section that the index of embedding
is a good indicator for the existence of $G/H$ instantons  in this case. 
Thus one should be 
able to make a connection between $\alpha$ and $\pi_3 (G/H)$. 
In fact, we find that in this case 
\beq
\label{pi3}
\pi_3 (G/H) =Z_{\alpha},
\eeq
which is in a complete agreement with our expectations. If $\alpha =1$ then
$\pi_3 (G/H)$ is trivial, and all $G$ instantons are mapped 
trivially to the $H$ instantons.
However, if $\alpha >1$, then there are $Z_{\alpha}$ instantons in $G/H$,
which are not contained in $H$, and their effect has to be added to the
low-energy theory. 

\begin{figure}
\PSbox{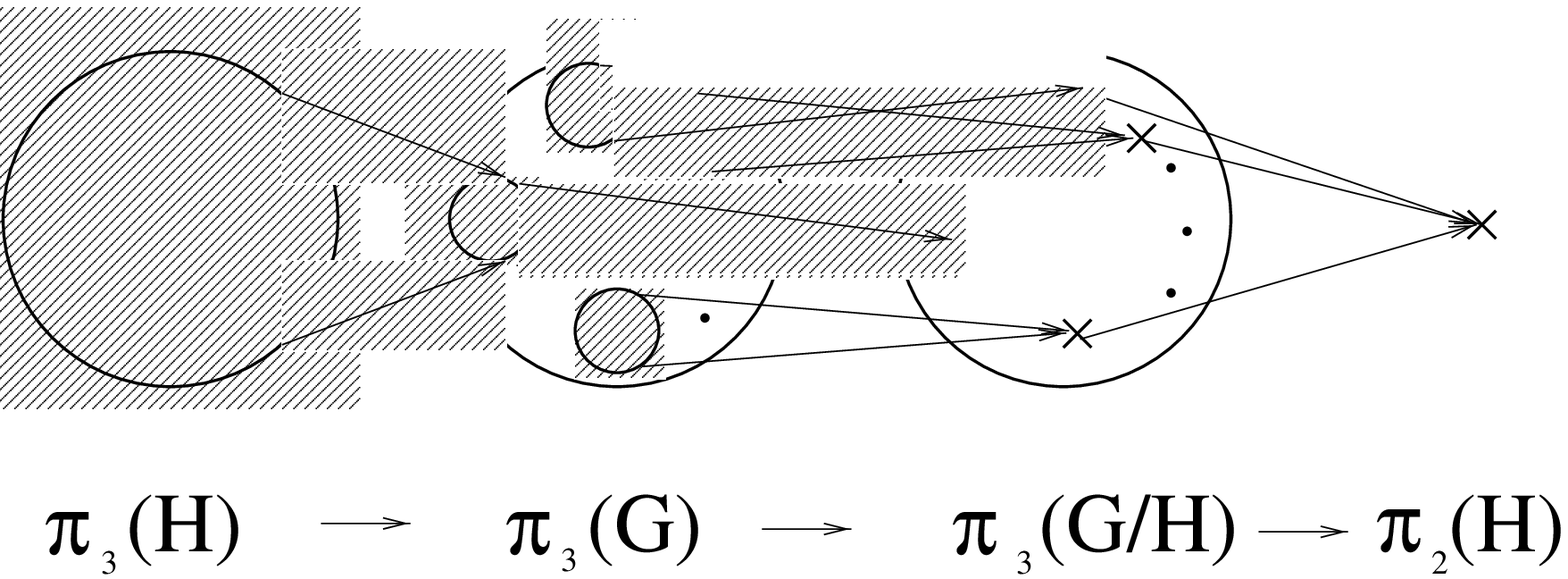 hscale=70 vscale=70 hoffset=50  voffset=0}{13.7cm}{4.5cm}
\caption{The exact homotopic sequence of (\protect\ref{exact}) for 
simple groups if the index of embedding is greater than one.
\label{sequence}}
\end{figure}

\begin{figure}
\PSbox{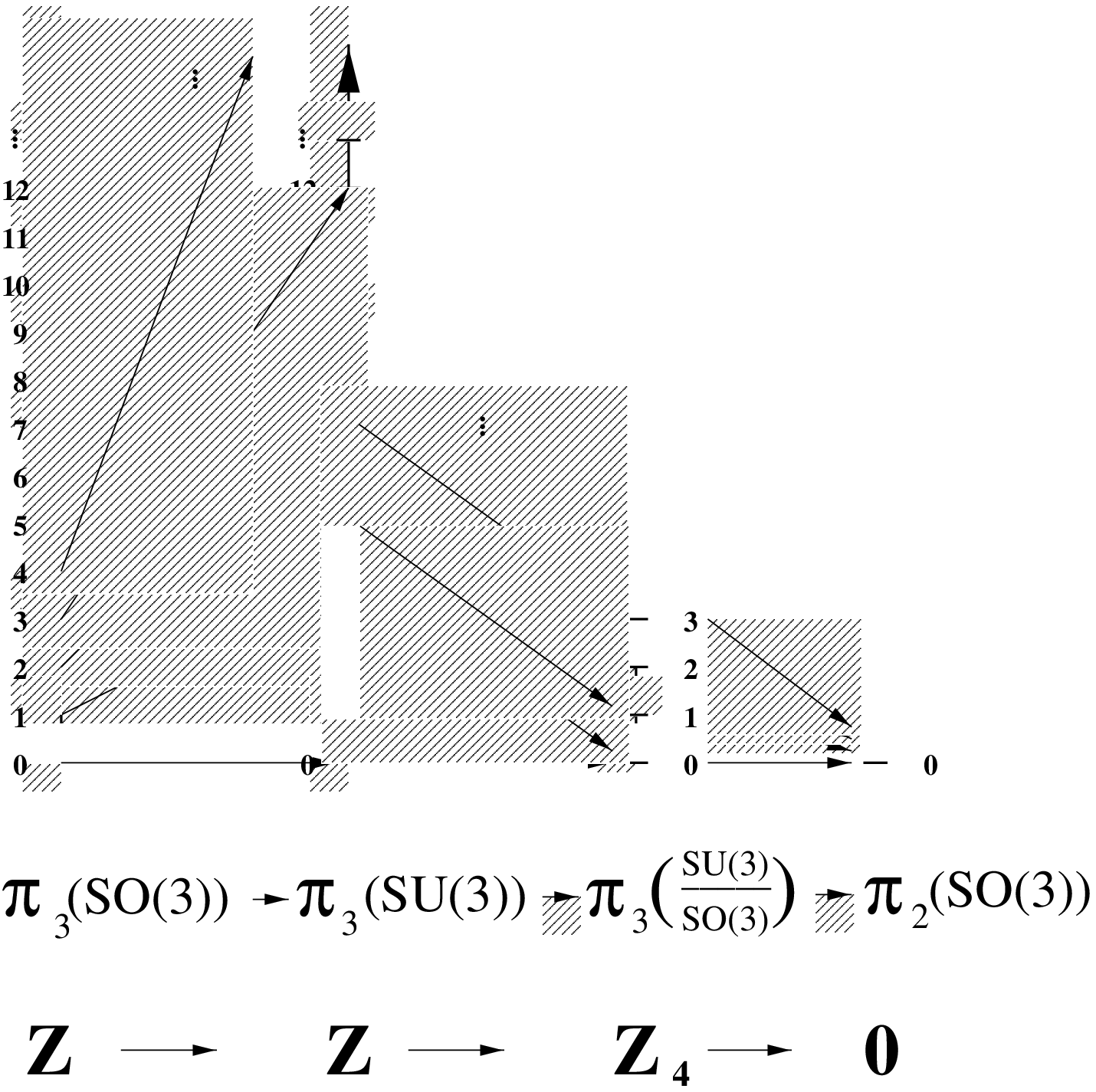 hscale=70 vscale=70 hoffset=100  voffset=0}{13.7cm}{8.5cm}
\caption{The exact homotopic sequence for the case that the index of embedding 
is four.\label{z4}}
\end{figure}

In order the establish the relation (\ref{pi3}) between the index of 
embedding and $\pi_3 (G/H)$, we consider the 
following part of the exact homotopic sequence:
\beq
\label{exact}
\ldots \to \pi_3 (H) \to \pi_3 (G) \to \pi_3 (G/H) \to \pi_2 (H) \to \ldots
\eeq
Since this sequence is exact, ${\rm Im} \, f_i={\rm Ker}
\, f_{i+1}$, where the $f$'s
denote the maps in (\ref{exact}). We know that $\pi_2 (H)=0$ for any
Lie groups, while with the assumption that $G$ and $H$ are
simple groups, $\pi_3 (G) = \pi_3 (H) = Z$. Thus we find that the sequence
\beq 
Z\to Z \to \pi_3 (G/H) \to 0
\eeq 
is exact. Since $\pi_2 (H)=0$ the kernel of the
map from $\pi_3 (G/H)$ to $\pi_2 (H)$ is the full $\pi_3 (G/H)$. Due
to the exact sequence, this means that the image of the map from
$\pi_3 (G)$ to $\pi_3 (G/H)$ is again the full $\pi_3 (G/H)$,
${\rm Im} (\pi_3 (G))=\pi_3 (G/H)$, where ${\rm Im} (\pi_3 (G))$ 
denotes the image of $\pi_3 (G)$. Therefore, $\pi_3 (G/H)= \pi_3 (G)/{\rm Ker} 
(\pi_3 (G))$, where ${\rm Ker}(\pi_3 (G))$ is the kernel of the map
from $\pi_3 (G)$ to $\pi_3 (G/H)$. 
However, due to the first part of the exact sequence 
${\rm Ker} (\pi_3 (G))={\rm Im} (\pi_3 (H))$. Thus
we can conclude that
\beq 
\pi_3 (G/H) = \pi_3 (G)/\mbox{Im}(\pi_3(H)),
\eeq
where $\mbox{Im}(\pi_3(H)) \subset \pi_{3}(G)$ is the image of $\pi_{3}(H)$.
Next we want to use the information that the index of embedding is $\alpha$ 
in order to relate $\pi_3 (G)$ and $\mbox{Im}(\pi_3 (H))$. We have seen in 
Section~\ref{index} that the fact that the index of embedding is 
$\alpha$ combined with the Atiyah-Singer index theorem implies that the
one instanton of $H$ corresponds to an $\alpha$ instanton 
of $G$. This means that winding around once in the $H$ subgroup
corresponds to winding around $\alpha$-times in the full $G$ group.
Since the value of $\pi_3$ measures how many times a given configuration 
is winding around the sphere at infinity, the above relation implies
that 
\beq
\label{winding}
\mbox{Im}(\pi_3(H))  = \alpha \times  \pi_3 (G),
\eeq
A more precise argument for  (\ref{winding}) is presented in Appendix
A. Combining the facts that 
\begin{eqnarray}
&& \pi_3 (G/H)=\pi_3(G)/\mbox{Im}(\pi_3(H)), \nonumber \\
&& \mbox{Im}(\pi_3(H))  = \alpha \times  \pi_3 (G), \nonumber \\
&&\pi_{3} (G) = Z
\end{eqnarray}
immediately gives the desired relation
\beq
\pi_3 (G/H)=Z_{\alpha}.
\eeq
Fig.~\ref{sequence} illustrates the exact sequence for this case when 
$G$ and $H$ are both simple.
Examples of non-trivial $\pi_3(G/H)$ include:
\begin{eqnarray}
&& \pi_3 (SU(N)/SO(N))=Z_2, \; \; (N>3) \nonumber \\
&& \pi_3 (SU(3)/SO(3))=Z_4, \nonumber \\
&& \pi_3 (Sp(2N)/SU(N))=Z_2.
\end{eqnarray}
The first two of these examples have been explicitly quoted in 
Ref.~\cite{Witten}. The case of the 
embedding of $SO(3)$ into $SU(3)$ (when the index $\alpha =4$) 
is illustrated in Fig.~\ref{z4}.\footnote{The $N=4$ case is somewhat 
special because $\pi_{3} (SO(4)) = Z\times Z$.  It is still true, 
however, that $\pi_3 (SU(4)/SO(4))=Z_2$ by following the same 
argument here for this particular case.}

The physical meaning of (\ref{pi3}) is
that during the breaking of $G$ to $H$ the instantons get 
separated into two categories. Some instantons remain in the unbroken
subgroup $H$, and are of the usual kind. However, there will be $Z_{\alpha}$
instantons in the partially broken group. Since these are $Z_{\alpha}$ and
not $Z$-type instantons, it means that a combination of $\alpha$ of these
instantons can unwind and be topologically trivial in $G/H$. This corresponds 
to the expectation that a collection of  $\alpha$ of $Z_{\alpha}$ instantons
will be ordinary instantons in $H$, and no longer in
$G/H$.\footnote{Note, however, that the $\alpha$-instanton configurations
  in $G$ are not exhausted by the one-instanton configurations in $H$
  because the former has much larger number of parameters.  Therefore
  $N\alpha$-instanton configurations, which belong to the
  topologically trivial homotopy class in $\pi_3 (G/H)$, have to be
  summed over in the dilute gas instanton summation.  This ensures
  the exponentiation of the 't Hooft operator so that it can be added
  to the action of the low-energy effective $H$ theory.}

In the case of $N=2$ theories, the low-energy $U(1)^r$ theory 
obtained after giving an expectation value to the adjoint
does not contain 
instantons any more, thus as explained at the end of Section~\ref{completely},
one has to add the instanton corrections to the low-energy
prepotential. This is expressed in the equation $\pi_3(G/U(1)^r)=Z$,
which tells us that all $G$ instantons are in the broken part of the 
gauge group, and thus their effects on the low-energy theory have to be added.

Let us now consider an example when $G$ is not a simple group.
This is the case for example in the breaking $SO(4)\to SO(3)$,
since $SO(4)\equiv (SU(2)\times SU(2))/Z_{2}$.\footnote{The $Z_{2}$ factor 
does not play any important role as long as $\pi_{3}$ is concerned.} 
To obtain $\pi_3 (SO(4)/SO(3))$ we note that $SO(4)/SO(3) \simeq S^3$,
which is a special case of the general relation
$SO(N)/SO(N-1)\simeq S^{N-1}$. 
Since $\pi_k (S^k)=Z$, we conclude that $\pi_3 (SO(4)/SO(3))=Z$.
This is again in accordance with our physics
expectations, since we have seen, that during the breaking of 
$SU(2)\times SU(2)$ to the diagonal $SU(2)$ one particular combination of the
one-instantons of the two $SU(2)$ factors (the $(1,1)$ instanton)
will be mapped to the 
one-instanton of the diagonal $SU(2)$. Thus the complete tower of the
other independent combination the instantons (the $(k,0)$ instantons
for example) are
missing from the diagonal $SU(2)$, and this is why $\pi_3 (G/H)=Z$ now.
Similarly, we find that for the general case of breaking
$SU(N)\times SU(N)$ to the diagonal $SU(N)$ subgroup, 
$\pi_3((SU(N)\times SU(N))/SU(N))=Z$. 
 
Thus we have seen that, during partial breaking  of the gauge group, 
some of the original $G$ instantons may get mapped to $Z_N$
instantons in $G/H$. The effects of these instantons are no longer included 
in the low-energy effective theory based on the gauge group $H$. However,
the effects of these $G/H$ instantons may leave non-trivial effects 
on the low-energy physics, and these have to be taken into account. In the
remainder of this paper we will show several examples of the effects 
of these $Z_N$ instantons on the low-energy physics. We will see that in
many cases, consistency of the low-energy theory will 
actually require the presence of  these $Z_N$ instanton effects.

\section{Examples of the Effects of $Z_N$ Instantons}
\setcounter{equation}{0}
\setcounter{footnote}{0}

In this section we present several examples of the effects of the 
$Z_N$ instantons discussed in the previous section on the low-energy
effective theory. We will focus on supersymmetric theories, since the
low-energy dynamics of these theories is much better understood than 
for general non-supersymmetric theories. Nevertheless, the arguments
of the previous section do apply to the non-supersymmetric theories as
well.

\subsection{The $Z_4$ Instantons in $SU(3)\to SO(3)$ \label{SU3}}

In this example we consider the $N=1$ duality of Pouliot and Strassler of 
$SO(8)$ with one spinor and $F$ vectors~\cite{PS}. This theory is dual
to $SU(F-4)$ with one symmetric tensor and $F$ antifundamental fields,
and some gauge singlets. The duality is described in Table~\ref{tab:duality},
where the superpotential of the dual $SU(F-4)$ theory is
\beq 
\label{dualsup}
W_{dual}=  \alpha NSq^2 +\beta T{\rm det}\, S,
\eeq
where $\alpha$, $\beta$ are coupling constants.
The operators are matched as $V^{i} V^{j} \leftrightarrow N^{ij}$, 
$p^{2} \leftrightarrow T$.
\begin{table}
\[
\begin{array}{c|cccc}
& SO(8) & SU(F) & U(1) & U(1)_R \\ \hline
V & 8_v & \Yfund & -(F-4) & 1-6/F \\
p & 8_s & 1 & F(F-4) & 1 \\ \hline \hline
& SU(F-4) & SU(F) & U(1) & U(1)_R \\ \hline
S & \Ysymm & 1 & -2F & 0 \\
q & \overline{\Yfund} & \overline{\Yfund} & 2F-4 & 6/F \\
N & 1 & \Ysymm & -2(F-4) & 2-12/F \\
T & 1 & 1 & 2F(F-4) & 2 \end{array},
\]
\caption{The field content and symmetries of the electric $SO(8)$ theory
and its magnetic dual $SU(F-4)$.\label{tab:duality}}
\end{table}
Integrating out the spinor $p$ of the electric theory with a mass 
term $\Delta W \propto p^{2}$ should reproduce 
the $SO$ duality of Ref.~\cite{SO}, since on the electric side we get 
an $SO(8)$ theory with $F$ vectors. On the magnetic side the mass of the spinor
corresponds to a linear term in $T$ in the superpotential, which will become
\beq 
\label{linearT}
W_{dual}= \alpha NSq^2 +\beta T{\rm det}\, S +\gamma T,
\eeq
with $\gamma$ a coupling constant.
The $T$ equation of motion forces an expectation value to ${\rm det}\, S$,
which breaks $SU(F-4)$ to $SO(F-4)$, while the $NSq^2$ term will turn into
the $Nq^2$ superpotential of the $SO$ duality~\cite{PS}. 
However, in the special case of $F=7$, the dual gauge group is $SO(3)$, and
an additional superpotential term ${\rm det}\, N$ is needed for the 
$SO$ duality of Ref.~\cite{SO}. This exactly happens when the breaking is
$SU(3)\to SO(3)$, that is when the index of embedding is four. We will
now show that the ${\rm det}\, N$ term in the superpotential required
for duality is indeed generated by a $Z_4$ instanton effect. 

\begin{figure}
\PSbox{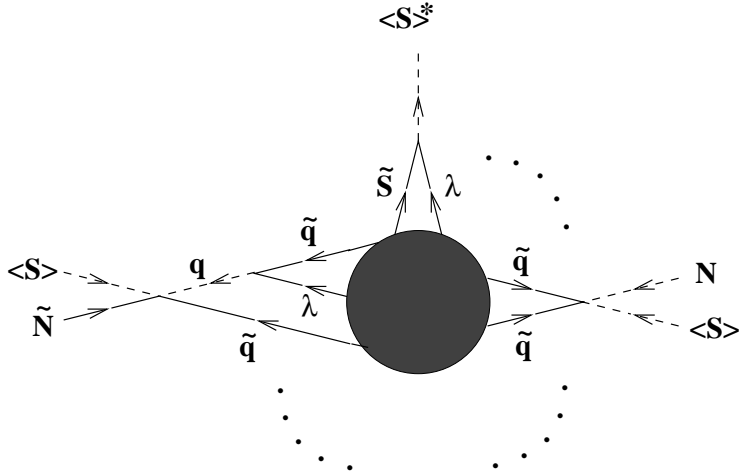 hscale=50 vscale=50 hoffset=80  voffset=0}{13.7cm}{6.5cm}
\caption{The contribution of the $Z_4$ instanton to the superpotential.
The blob in the middle represents the two-instanton of $SU(3)$, which 
is one of the $Z_4$ instantons. The straight lines represent fermions
while the dashed ones scalars. The fermionic lines emerging from the
instanton form the 't Hooft vertex. In addition, as explained in the text,
several insertions of the scalar-fermion-gluino vertex and vertices
from the superpotential $Nsq^2$ are needed to convert the 't Hooft vertex
into a superpotential contribution.\label{fig:instanton}}
\end{figure}

For this we consider the two-instanton of $SU(3)$. This is 
one of the three instantons which is missing from the $SO(3)$ theory.
The 't Hooft effective Lagrangian for this two-instanton is given by
\beq
\label{tHooft}
\tilde{S}^{10} \tilde{q}^{14}\lambda^{12} \Lambda^{6},
\eeq
where $\tilde{S}$ is the fermionic component of the chiral superfield
$S$,  $\tilde{q}$ the fermion from $q$, $\lambda$ the gaugino, 
and $\Lambda^{6}$ is the two-instanton factor.
The powers of these fields are fixed by the number of zero modes in the
two-instanton background. In order to show that this will indeed result in
a superpotential of the form ${\rm det} N$, we need to convert 
the fields in (\ref{tHooft}) to $\tilde{N}^2 N^5$ for the case of $F=7$, since
this is one of the contributions of the superpotential to the 
Lagrangian. Here $\tilde{N}$ is the fermionic component of $N$. 
In the presence of the expectation value
$\langle S \rangle \propto (\gamma/\beta)^{1/3}$,
the gaugino vertex $S^{*} \lambda \tilde{S}$ converts $\tilde{S}^{10} 
\lambda^{10}$ to ${\langle S\rangle^*}^{10}$. The integration over the 
instanton size will result in additional factors of $\langle S \rangle
\langle  S \rangle^*$.
Since we are interested in a contribution to the
superpotential, all dependence on $\langle S\rangle^*$ has to cancel 
after the integral over the instanton size is performed, due to the 
holomorphy of the superpotential. This can happen only if for every factor
of $\langle S\rangle^*$ there is a $\langle S \rangle\langle  S \rangle^*$
dividing the operator. Thus, every factor of $\langle S \rangle^*$ is
converted to $\langle S\rangle^{-1}$, and so  ${\langle S\rangle^*}^{10}$
has to be replaced by $\langle S\rangle^{-10}$ for the 
holomorphic part which can appear in the superpotential. 
The superpotential 
coupling $\alpha N S q^{2}$ converts ten  $\tilde{q}$'s
out of $\tilde{q}^{14}$ to $(\alpha N 
\langle S \rangle)^{5}$ by using the vertex $\alpha N S \tilde{q}^2$ 
five times, and the other four $\tilde{q}$'s 
together with the remaining two
$\lambda$'s to the fermionic component of 
$(\alpha N \langle S\rangle)$ by using two $\tilde{q}\lambda q^*$ and two
$\alpha S q \tilde{N}\tilde{q}$ vertices. This is illustrated in 
Fig.~\ref{fig:instanton}. In total, the operator generated can be 
written in terms of the superpotential term
\begin{equation}
	\frac{\mbox{det}\, (\alpha N \langle S\rangle) \Lambda^{6}}
		{\langle S\rangle^{10}} .
\end{equation}
This form is consistent with all symmetries of the theory and has the 
right dimensionality to be a term in the superpotential.
Up to a dimensionless constant $(\alpha \langle S \rangle)^{7}$, 
the superpotential can be rewritten as
\beq 
\label{halfinst}
\frac{{\rm det}\, N\, \Lambda^{6}}{\langle S\rangle^{10}} = 
\frac{{\rm det}\, N}{\tilde{\Lambda}^4},
\eeq
where $\tilde{\Lambda}$ is the scale (Landau pole) of the $SO(3)$ 
theory.  However, the one-loop $\beta$ function of the $SO(3)$ theory 
is $-8$, therefore the expression of (\ref{halfinst}) corresponds as 
expected to a ``half-instanton'' effect in the $SO(3)$ group, which 
can only be explained as a $Z_4$ instanton effect in the partially 
broken group.

Thus one can see that the effect of one of the $Z_4$ instantons is to
generate the superpotential term ${\rm det} N$.  However, one needs to ask 
the question of why we took only the effect of the two instanton into account,
and not those of the one and three instantons which are not present in the
low-energy theory either.  The absence of the effects of these instantons 
in the superpotential can be understood by considering the global charges
of the theory.\footnote{The following arguments do not exclude the possibility
that the one-instanton configuration generates an irrelevant operator 
in the K\"ahler potential.}  
 Consider the anomalous $R$-symmetry $U(1)_X$, under which
$S$ and $q$ have charge zero (the fermionic components have charge
$-1$), and the $SU(3)$ gauginos have charge $1$. In order for the 
superpotential (\ref{dualsup}) to carry charge $2$, $N$ and $T$ have to have
$U(1)_X$ charge $2$. In addition, the non-anomalous $R$-charges can be read
off from Table~\ref{tab:duality} for $F=7$. Thus the $R$-charges under
these two symmetries are:
\[
\begin{array}{|c|cccc|} \hline
& S & q & N & T \\ \hline
U(1)_R & 0 & 6/7 & 2/7 & 2 \\
U(1)_X & 0 & 0 & 2 & 2 \\ \hline \end{array} \]
Note that both the $U(1)_{R}$ and $U(1)_{X}$ remain symmetries of the 
model at 
the classical level even with the added linear term $T$ in the superpotential.
The 't Hooft vertex of the one-instanton of $SU(3)$ is $\lambda^6 \tilde{S}^5
\tilde{q}^7$ which carries $U(1)_X$ charge $-6$ (and of course $U(1)_R$ charge 
zero). If this vertex is to come from a superpotential, then that 
superpotential term has to carry $U(1)_X$ charge $-4$ (and $U(1)_R$ charge
$2$). Thus the difference between the $R$ and $X$ charge must be $6$. 
Below is a list of gauge and global invariants of the theory, of which the
superpotential term must be constructed from:
\[ 
\begin{array}{|c|ccc|} \hline
& U(1)_R & U(1)_X & R-X \\ \hline
{\rm det} S & 0 & 0 & 0 \\
T & 2 & 2 & 0 \\
{\rm det} (Sq^2) & 12 & 0 & 12 \\
Nsq^2 & 2 & 2 & 0 \\
{\rm det} N & 2 & 14 & 12 \\ \hline \end{array} \]
One can see that the difference $R-X$ is either $\pm 12$ or $0$, thus $6$ 
can not be obtained in any way, therefore the 1-instantons (and similarly 
the 3-instantons) can not generate a superpotential term, but the 
2-instanton can, as we have explicitly seen above.

In fact, the absence of the one-instanton effect is expected from the 
$SO(N)$ duality of the low-energy theories.  Based on duality of the 
theories obtained after adding the spinor mass, we show below not 
only that there should not be any superpotential terms generated by the 
one (or three) instantons, but also that all effects of these instantons must 
decouple completely from the low-energy theory.  This can be seen for 
arbitrary $F$ by considering the discrete symmetries of the theories.  
The original $SO(8)$ electric theory does not have a non-trivial 
discrete symmetry not contained in continuous symmetries, nor the 
magnetic $SU(3)$ theory.  However after the spinor mass term is added, 
the electric theory flows to the $SO(8)$ theory with $F$ vectors, 
which has a $Z_{2F}\times {\cal P}$ discrete symmetry, where $Z_{2F}$ 
acts as $Q\to e^{i\pi /F}Q$ on the $SO(8)$ vectors, and ${\cal P}$ is 
the color conjugation type discrete symmetry, under which the sign of 
the first color is flipped (color parity)~\cite{discrete}.  If duality 
is to hold, the low-energy $SO(3)$ theory has to have the same set of 
discrete symmetries.  One discrete symmetry of the dual $SO(F-4)$ 
theory is obtained as the unbroken discrete subgroup of the original 
symmetries of the electric theory.  Adding the term linear in $T$ to 
the superpotential breaks the ordinary global $U(1)$ to its 
$Z_{2F(F-4)}$ subgroup, under which the charges of $S,q,N$ and $T$ are 
$-2F,2F-4,-2(F-4)$ and $0$, respectively.  However, the equation of 
motion for $T$ forces an expectation value to $S$, which would break 
this $Z_{2F(F-4)}$ symmetry further.  To find the unbroken discrete 
symmetry, let us combine the action of the above $Z_{2F(F-4)}$ with a 
global $SU(F-4)$ gauge transformation $U$ of the form \[U= {\rm diag} 
(-e^{\pi i/(F-4)}, e^{\pi i/(F-4)}, \ldots , e^{\pi i/(F-4)}).  \] 
This is an element of $SU(F-4)$ since the determinant is one.  Acting 
by this $U$ on $S$ as $S\to U^T S U$, and combining this with the 
action of the $Z_{2F(F-4)}$, $\langle S\rangle$ is left invariant.  
Thus this is an unbroken discrete global symmetry of the theory.  Now 
let us determine how this symmetry acts on the $q$'s.  The gauge 
transformation acts as $q\to U^{\dagger} q$, since $q$ is in a 
representation conjugate to $S$, while $q$ has charge $2F-4$ under the 
$Z_{2F(F-4)}$.  Thus the action of this symmetry is given by \[ q\to 
{\rm diag} (-e^{-\pi i/(F-4)}, e^{-\pi i/(F-4)}, \ldots , e^{-\pi 
i/(F-4)}) e^{\frac{\pi i(2F-4)}{F(F-4)}} q,\] which is just \[ 
(q_1,q_2,\ldots ,q_{F-4}) \to e^{\pi i/F} (-q_1, q_2,\ldots 
,q_{F-4}).\] One can see that this $Z_{2F}$ discrete symmetry is 
nothing but the combination of the $Z_{2F}$ symmetry which acts as 
$q\to e^{\pi i/F}q$ with the color-parity transformation ${\cal P}$.  
This is the discrete symmetry in the magnetic theory which is mapped 
to the $Z_{2F}$ symmetry of the electric theory.  However, the 
color-parity ${\cal P}$ itself does not arise from the original 
symmetries of the $SU(F-4)$ theory, but it is an accidental symmetry 
of the low-energy effective theory.  The 't Hooft one-instanton vertex 
is invariant under the combination of $Z_{2F}{\cal P}$, since it is 
invariant under every global and gauge symmetry of the theory.  
However, as explained above ${\cal P}$ is not part of the symmetries 
of the original $SU(F-4)$ theory, and the 't Hooft one-instanton 
vertex is not necessarily invariant under it.  Indeed, since the 
one-instanton vertex contains exactly one $SU(F-4)$ epsilon-tensor, it 
changes sign under ${\cal P}$ and is thus not invariant, and the 
effects of this vertex would violate the ${\cal P}$ symmetry.  However, 
by duality we expect that ${\cal P}$ itself is a good symmetry of the 
low-energy effective $SO(F-4)$ theory, therefore all effects of the 
one-instanton (which would break this symmetry) must decouple from the 
low-energy effective theory.

\subsection{The $Z_2$ Instanton in $N=2$ $Sp(2N)\to SU(N)$ \label{N=2Sp}}

In this example we consider pure $N=2$ $Sp(2N)$ theories. This theory is in
the Coulomb phase, and the low-energy effective action can be obtained from 
the following hyperelliptic Seiberg-Witten curve~\cite{Sp}:
\beq
\label{Spcurve}
y^2=x\prod_{i=1}^N (x-\Phi_i^2)^2+4x\prod_{i=1}^N (x-\Phi_i^2)
\Lambda_{Sp}^{2N+2},
\eeq
where $x$ and $y$ are the coordinates of the Seiberg-Witten curve, 
$\Lambda_{Sp}$ is the dynamical scale of the $Sp(2N)$ theory and 
$\Phi_i$ are the eigenvalues of the adjoint (symmetric tensor) of the
$Sp(2N)$. We will show that by higgsing to the $SU(N)$ subgroup one 
obtains a curve that is different from the usual $SU(N)$ Seiberg-Witten 
curve. We will see explicitly that the effect of the $Z_2$ instanton 
is a shift in the Seiberg-Witten curve which modifies the singular locus
of the curve.

Let us consider the breaking of the $Sp(2N)$ theory to its $U(N)$ subgroup.
This is achieved by giving an expectation value $\Phi_i=V$ to the
adjoint of $Sp(2N)$. This embedding has index two, thus
there are potential effects of the $Z_2$ instanton on the low-energy 
effective theory. Writing $\Phi_i=V+\tilde{\Phi}_i/2$ in  (\ref{Spcurve})
and redefining $x$ as $x-V^2=Vx'$ we get
\beq
y^2=(Vx'+V^2)^2\prod_{i=1}^N \left[ V^2 (x'-\tilde{\Phi}_i-
\frac{\tilde{\Phi}_i^2}{4V})^2 \right]+4(Vx'+V^2)\Lambda_{Sp}^{2N+2}
\prod_{i=1}^N \left[ V(x'-\tilde{\Phi}_i-
\frac{\tilde{\Phi}_i^2}{4V}) \right]. \nonumber \\
\eeq 
Taking the $V\to \infty$ limit and dropping the prime from $x$ 
and the tilde from $\Phi$, we obtain 
the curve 
\beq
y^2=V^{2N+4}\prod_{i=1}^N (x-\Phi_i)^2+4V^{N+2}\Lambda^{2N+2}_{Sp}
\prod_{i=1}^N (x-\Phi_i).
\eeq
The scale matching relation according to (\ref{matching}) is given by
\beq 
(\Lambda_{Sp}^{2N+2})^2=\Lambda_{SU}^{2N}V^{2N+4}.
\eeq
Thus after rescaling $y$, we finally obtain the following curve:
\beq 
\label{finalcurve}
y^2=\left( \prod_{i=1}^N (x-\Phi_i)^2+2\Lambda_{SU}^N\right)^2
-4\Lambda_{SU}^{2N}.
\eeq
This results needs more explanation. We can see that, except for the
shift of the gauge-invariant polynomial $\prod_{i=1}^{N} \Phi_{i}$ by 
$2\Lambda_{SU}^N$, we have obtained the usual Seiberg-Witten curve 
for pure $N=2$ $SU(N)$ theories.\footnote{Since no field is charged under 
the $U(1)$ part of the $U(N)$ gauge group this $U(1)$ decouples from the
low-energy theory.} This shift is proportional to $\Lambda_{SU}^N$, which 
is the square root of the one instanton factor of $SU(N)$, therefore can not
be due to an instanton effect in the $SU(N)$ theory. Instead, this shift of
the curve should be interpreted as the effect of the $Z_2$ instanton on the
low-energy effective theory. Thus even in the $V\to \infty$ limit the
effects of the $Z_2$ instanton do not decouple from the low-energy theory   
effective theory; it ``remembers'' that it has been obtained by higgsing 
from the $Sp(2N)$ theory. One can easily see that the two curves are not 
equivalent by comparing the discriminant of the curve in (\ref{finalcurve}) 
to the discriminant of the usual $SU(N)$ curve. For example,
the curve of (\ref{finalcurve}) obtained by higgsing $Sp(4)$ to 
$SU(2)$ is  given explicitly by
\beq
\label{curve}
y^2=\left( x^2-u-2\Lambda_{SU}^2 \right)^2 -4\Lambda_{SU}^4,
\eeq
while the curve for the pure $N=2$ $SU(2)$ theory is given by
\beq
\label{SU2curve}
y^2=(x^2-u)^2 -4\Lambda_{SU}^4.
\eeq
In the usual $SU(2)$ theory the singularities
occur at $u=\pm 2 \Lambda_{SU}^2$, 
in the theory given by the curve of (\ref{curve})
the singularities occur at $u=0, -4\Lambda_{SU}^2$. One may ask the question
of whether this shift in the curve (and in the position of the singularity) is
a physically observable effect. From the purely low-energy point of
view one could argue that this shift just amounts to a redefinition of the
coordinates on the moduli space, and therefore in the strict low-energy 
limit this effect is unobservable. 
However, if one considers not only the
low-energy effective theory but also the high-energy theories, 
effect of the shift is actually physically observable. 
One way of seeing this is to
remember that the $SU(N)$ theory has a non-anomalous $Z_{2N}$ 
discrete symmetry, under which the adjoint field has charge one,
thus $u$ of the above example carries charge two. We have seen that
the singularity occurs at non-zero values of $u$ in the pure $N=2$ $SU(2)$ theory, 
thus breaking the discrete symmetry. However, in the $SU(2)$
theory obtained from higgsing the $Sp(4)$ the effects of the $Z_2$ instanton
shift one of the singularities to zero. Thus in this effective $SU(2)$
theory the discrete symmetry is not broken at one of the singularities. 
The coexistence of the unbroken discrete symmetry and the massless monopole
is the effect of the $G/H$ instantons.

One can discuss the same issue from a different point of view.
Consider 
the  pure $N=2$ $SU(2)$ theory as a high-energy theory. This
theory has an anomalous $U(1)_R$ symmetry, which can be used to obtain
selection rules if one assigns an appropriate charge to the
dynamical scale $\Lambda_{SU}^4$ of the theory.  Requiring invariance under 
this symmetry will tell us what kind of redefinitions of the moduli
space parameter $u$ are possible. Since the shift in $u$ 
required to connect the two curves in (\ref{curve}) and (\ref{SU2curve})
would be proportional 
to the square root of the instanton factor $\Lambda_{SU}^4$, this
would signal that the $U(1)_R$ symmetry is broken, therefore such a shift is
not allowed by the anomalous $U(1)_R$ symmetry. Thus taking into account the
symmetries of the high-energy theory distinguishes between the two curves
presented above. We will see in the following section the same effect
again in the case of $SO(N)$ theory, where the
shift in the curve changes the fact whether some non-anomalous 
continuous global 
symmetries of the high-energy theory are preserved at the singularity or not.

\subsection{The $Z$ Instantons in $SU(N)\times SU(N)\to SU(N)_D$}

In this section we will consider examples very similar to the $N=2$ theory
presented in the previous section. Here we consider $N=1$ product group 
theories in the Coulomb phase~\cite{phases,JoshDan}. Let us consider first the
$SU(2)\times SU(2)$ model of Ref.~\cite{phases}. The matter content
is given by
\[
\begin{array}{c|cc}
& SU(2) & SU(2)\\ \hline
Q_1 & \Yfund & \Yfund \\
Q_2 & \Yfund & \Yfund \end{array}. \]
This theory is in the Coulomb phase, and the Seiberg-Witten curve is given 
by
\beq 
y^2=(x^2-(u-\Lambda_1^4-\Lambda_2^4))^2-4\Lambda_1^4\Lambda_2^4,
\eeq
where $u={\rm det}\, M$, $M_{ij}=Q_iQ_j$, and the $\Lambda_{1,2}$ are the
dynamical scales of the two $SU(2)$ factors. Let us give an expectation value 
to $Q_1$, thereby breaking $SU(2)\times SU(2)$ to the diagonal $SU(2)$ 
subgroup. We have seen in Section~\ref{sec:pi3} that there are potential
$G/H$ instantons appearing in this embedding which might have an effect on the
low-energy theory. To find their effect on the low-energy theory
we write $M_{11}=V^2$, and the scale matching relation 
$\Lambda_D^4=\frac{\Lambda_1^4\Lambda_2^4}{V^4}$. Denote the
ratio of the two $SU(2)$ scales $\frac{\Lambda_1^4}{\Lambda_2^4}=k^2$.
Then the low-energy curve can be written as
\beq
y^2=(x^2-V^2(u_D-(k+k^{-1})\Lambda_D^2))^2-4\Lambda_D^4V^4.
\eeq
After an appropriate rescaling of $x$ and $y$ the curve becomes
\beq 
\label{IntSeibcurve}
y^2=(x^2-(u_D-(k+k^{-1})\Lambda_D^2))^2-4\Lambda_D^4.
\eeq
The conclusion is similar as in the previous example. The Seiberg-Witten 
curve differs from the ordinary $SU(2)$ curve, and contains effects which can 
not be explained by instanton effects in the diagonal $SU(2)$ theory. Instead
they are due to the $G/H$ instantons. Note that here the low-energy
theory depends also on the ratio of the scales of the original two 
$SU(2)$ groups, which is not lost in the effective $SU(2)$ curve exactly due 
to the effects of the $G/H$ instantons. Note that for $k=1$, that is in the
case when the two scales of the original $SU(2)$ theories coincide, 
the location of one of the singularities is shifted to the origin of the
moduli space. This changes the monodromies, and implies that additional
monopoles must become massless at this point. Thus one can see that the
shift in the curve, which is an effect of the instantons in the
broken part of the gauge group, encodes important physical 
information. 

It is straightforward to generalize this example to the general
$SU(N)\times SU(N)$ theories of Ref.~\cite{JoshDan}, which we will 
briefly review at the end of this section.
First, however, let us examine the $SO(N)$ theory with $N-2$ vectors 
discussed in Ref.~\cite{SO}, which is very closely related to the
$SU(2)\times SU(2)$ theory analyzed above. 
This theory is again in the Coulomb phase,
and the Seiberg-Witten curve is given by
\beq
\label{SOcurve}
y^2=\left[ x^2-(U-2\Lambda_{SO}^{2N-4})\right] -4\Lambda_{SO}^{4N-8},
\eeq
where $U={\rm det}\, M$, $M$ is the meson matrix $M_{ij}=Q_iQ_j$, $i=1,\ldots ,
N-2$, and $\Lambda_{SO}$ is the dynamical scale of the $SO(N)$ theory.
At the origin of the moduli space $M_{ij}=0$ the $SU(N-2)$ global 
symmetry arising from rotations of the $N-2$ vectors is unbroken, and the
't Hooft anomaly matching conditions have to be satisfied. One finds that 
this is indeed the case, once we realize that the curve (\ref{SOcurve})
has a singularity at the origin, and $N-2$ monopoles transforming as an
antifundamental representation under the $SU(N-2)$ global symmetry become
massless. Thus one can see that the fact that one of the singularities is
precisely at the origin plays a crucial physical role. Let us now
examine how this singularity at the origin arises. In order to obtain the
curve of (\ref{SOcurve}) one breaks the $SO(N)$ theory to an 
$SO(4)\sim SU(2)\times SU(2)$ theory by giving an expectation value
to $N-4$ vectors. This way one obtains an $SU(2)\times SU(2)$ theory 
with exactly the same matter content as discussed above, and the
two scales equal, thus $k=1$. Further breaking to the diagonal $SU(2)$ 
subgroup as discussed above will determine the curve (\ref{SOcurve})
uniquely. The shift of the $SU(2)$ curve due to the effects of the 
$G/H$ instantons will result in the shift $2\Lambda_{SO}^{2N-4}$ in the
curve for the $SO(N)$ theory. This shift, as explained above, is crucial
for the 't Hooft anomaly matching, and is, in the low-energy $SU(2)$ 
theory, due to the effects of the $G/H$ instantons.

We close this section by explaining how to generalize the results for
the $SU(2)\times SU(2)$ theory presented at the beginning of this section
to $SU(N)\times SU(N)$. The matter content of the theories we consider 
is given by
\[
\begin{array}{c|cc}
& SU(N) & SU(N) \\ \hline
Q_1 & \Yfund & \overline{\Yfund} \\
Q_2 & \overline{\Yfund}  & \Yfund 
\end{array}. \]
This theory is in the Coulomb phase, with $N-1$ unbroken $U(1)$ factors at the
generic point of the moduli space. The Seiberg-Witten curve for this
theory has been determined in Ref.~\cite{JoshDan}. The independent gauge 
invariant operators are
$B_1={\rm det} Q_1$, $B_2={\rm det} Q_2$, $T_n={\rm Tr} (Q_1Q_2)^n$, 
$n=1,\ldots , N-1$. The Seiberg-Witten curve for this theory is
\beq
y^2=
\left( \sum_{i=0}^N s_ix^{N-i}+(-1)^N (\Lambda_1^{2N}+\Lambda_2^{2N})
\right)^2 -4\Lambda_1^{2N}\Lambda_2^{2N},
\label{prodcurve}
\eeq
where the $s_i$ are related to the $u_k$'s by Newton's formula
\beq
ks_k+\sum_{j=1}^kjs_{k-j}u_j=0,
\eeq
$s_0=1,s_1=u_1=0$, and the operators $u_k$ are the invariants
of the ``composite adjoint'' $\Phi =Q_1Q_2-\frac{1}{N} {\rm Tr} Q_1Q_2$,
$u_k=\frac{1}{k}{\rm Tr} \Phi^k$, and are to be expressed in terms of the 
gauge invariants $T_i$ and $B_i$ via classical expressions (for example
for the case of $SU(3)\times SU(3)$ $u_2=\frac{1}{2} (T_2-\frac{1}{3}T_1^2),
u_3=\frac{1}{3} (3B_1B_2+\frac{1}{2}T_2T_1-\frac{5}{18}T_1^3)$).

Now consider breaking $SU(N)\times SU(N)$ to the diagonal 
$SU(N)_D$ subgroup, by the expectation value
\beq
Q_1 = v \left( \begin{array}{cccc}
1\\&1\\&&\ddots\\&&&1 \end{array} \right),
\eeq
that is by giving a VEV to $B_1$ and no other operator. The matching of 
scales is given by $\frac{\Lambda_1^{2N}\Lambda_2^{2N}}{v^{2N}}=
\Lambda_D^{2N}$, while the operators $u_k$ will be matched to the invariants
of the adjoint of the diagonal $SU(N)_D$ by $u_k^D=\frac{u_k}{v^k}$.
Plugging these relations back into (\ref{prodcurve}), and rescaling
$x\to x/v$, $y\to y/v^N$, we obtain the curve for the $SU(N)_D$ theory:
\beq
y^2=\left( \sum_{i=0}^N s_i^Dx^{N-i}+(-1)^N(k+\frac{1}{k})\Lambda_D^N
\right)^2 -4\Lambda_D^{2N},
\eeq
where $k=\frac{\Lambda_1^{N}}{\Lambda_2^N}$, and $s_i^D$ are the symmetric
variables for the diagonal group $s_i^D=\frac{s_i}{v^i}$. The
conclusion is just like before: the Seiberg-Witten curve we obtain in this
limit is almost identical to the usual $N=2$ Seiberg-Witten curve, but
it differs from it by a shift due to the $G/H$ instantons in the broken part
of the group. Again the relative sizes of the original
scales $\Lambda_1$ and
$\Lambda_2$ appear in the low-energy theory due to the $G/H$ instanton
effects.

\subsection{The $Z$ Instantons in $SO(4)\to SO(3)$}

In this example we consider the breaking $SO(4)\to SO(3)$ by looking at 
the $N=1$ duality in $SO(N)$ groups with vectors discussed in Ref.~\cite{SO}.
The electric theory is
\begin{equation}
\label{SOtable}
\begin{array}{c|cccc}
& SO(N)&SU(F)&U(1)_R & Z_{2F} \\ \hline
Q & \Yfund & \Yfund & 1-\frac{N-2}{F} & 1 \end{array},
\end{equation}
while the dual is $SO(F-N+4)$ with $F$ vectors:
\begin{equation} \begin{array}{c|cccc}
& SO(F-N+4) & SU(F) & U(1)_R & Z_{2F} \\ \hline
q & \Yfund & \overline{\Yfund} & \frac{N-2}{F} & -1 \\
M & 1 & \Ysymm & 2-2\frac{N-2}{F} & 2 \end{array}, \end{equation}
 and a superpotential $Mq^2$, where the $q$'s are the magnetic quarks 
and $M$ are the mesons. However, in the case of $F=N-1$, the dual 
is $SO(3)$ with $F$ vectors, but the superpotential includes an 
additional term $W=Mq^2+{\rm det} M$. This ${\rm det} M$ term is present in 
the dual superpotential only for $F=N-1$. This prompts the question of how this
${\rm det} M$ term is generated if one starts from the duality for 
$F=N$ and integrate out one flavor. Thus we consider $SO(N)$ with $N$ vectors
and no superpotential. The dual is $SO(4)$ with $F$ vectors $q$, a gauge
singlet meson $M$ and a superpotential $Mq^2$. Adding a mass term for one 
vector of the electric theory results in an $SO(N)$ electric theory
with $N-1$ vectors. On the dual side the mass term corresponds to adding 
a term linear in the meson field to the superpotential. Thus the
full superpotential is $W=Mq^2+mM_{N,N}$. The equation of motion 
with respect to $M_{N,N}$ forces an expectation value to one of the
dual quarks higgsing the dual gauge group from $SO(4)$ to $SO(3)$.
Thus we get the non-trivial embedding of $SO(3)$ into $SO(4)$, that is
$SU(2)_D$ into $SU(2)\times SU(2)$. The effect of this is that some of the
instantons which are in the broken part of the group are no longer 
included in the low-energy theory. The effects of these instantons will be
exactly to reproduce the superpotential term ${\rm det} M$ required by 
duality. 

We describe how this term is generated by
the instantons in the broken group. 
The $(1,0)$ instanton configuration in the first $SU(2)$ factor of
$SO(4) \simeq (SU(2)_{L} \times SU(2)_{R})/Z_{2}$ 
generates the 't Hooft operator
\begin{equation}
\label{tHooft2}
	\tilde{q}^{1} \tilde{q}^{1} \cdots \tilde{q}^{N} \tilde{q}^{N}
        \lambda^4 \Lambda_{L}^{6-2N},
\end{equation}
where $\lambda$ is the gaugino in the first $SU(2)$ factor and 
$\Lambda_{L}$ is the scale of the first $SU(2)$ factor.
In the presence of the expectation value $\langle q^{N} \rangle \neq 
0$, both $\tilde{q}^{N}$'s in (\ref{tHooft2})
are contracted with the gauginos $\lambda$ to 
become ${q^{N}}^*$.  As explained in Section~\ref{SU3},
after the integral over the instanton size, 
a factor of $1/|\langle q^{N} \rangle|^{4}$ appears and the 
dependence on ${q^{N}}^*$ is canceled, and a factor of $1/\langle q^{N}
\rangle^{2}$ remains.  All the other $\tilde{q}^{i}$'s are contracted 
using the superpotential coupling $\hat{M}_{ij} q^{i} q^{j}$, 
where $\hat{M}$ is the submatrix of $M$ with $N$-th row and column 
removed, and become 
$\mbox{det}\, \hat{M}$ (four of them are combined with the remaining two
$\lambda$'s to give fermionic component of $\hat{M}_{ij}$).  The end result 
is the superpotential
\begin{equation}
\frac{(\mbox{det}\, \hat{M}) \Lambda_{L}^{6-N}}{\langle q^{N} \rangle^{2}},
\end{equation}
which is the term $\mbox{det}\, \hat{M}$ required for duality.   
Exactly the same superpotential is generated from the $(0,1)$ instanton 
configuration except for the replacement of $\Lambda_{L}$ by 
$\Lambda_{R}$.

An alternative way of obtaining the same superpotential term 
is to reduce the problem of that of $SU(2)\times SU(2)$ with one
representation in $(\Yfund ,\Yfund )$. The instanton corrections 
in this theory have been analyzed in Ref.~\cite{ILS}, therefore 
this method will result in the superpotential term including the 
appropriate coefficient. Below we briefly repeat this 
argument of Ref.~\cite{SO} as well. 
Consider the point on the moduli space of the dual $SO(4)$ theory where 
the meson has an expectation value of rank $N-1$. This gives mass to
all but one of the dual quarks $q$. On this point instantons (which will
later exactly correspond to instantons in the broken part of the $SO(4)$)
generate a superpotential
\beq
\label{diaginst}
W_{inst}=2\frac{\tilde{\Lambda}_L^5+ \tilde{\Lambda}_R^5}{q^Nq^N},
\eeq
where $q^N$ is the only massless flavor and the $\tilde{\Lambda}_{L,R}$ 
are the scales of the effective $SU(2)_L\times SU(2)_R$ theory after 
the $N-1$ flavors have been integrated out. The matching of scales
relates them to the original scale of the $SO(4)$ theory by
$\tilde{\Lambda}_{L,R}^5\propto {\rm det}\hat{M} \Lambda_{L,R}^{6-N}$.
After the mass to the last flavor has been added, $q^N$ will get an
expectation value, and the due to the matching relation the
superpotential of (\ref{diaginst}) will be exactly the ${\rm det} M$
term required by duality.

\subsection{The $Z_2$ Instanton in $N=1$ $Sp(2N)\to SU(N)$ \label{Sp}}
In this example we will consider $N=1$ supersymmetric
$Sp(2N)$ theories with a symmetric tensor (adjoint) and $2F$ fields
in the fundamental representation, and a tree-level superpotential
$W={\rm Tr}\, X^{2(k+1)}$. A duality for this theory has been 
described in Ref.~\cite{Spduality} and is summarized in the table below.
\[
\begin{array}{c|ccc}
& Sp(2N) & SU(2F) & U(1)_R \\ \hline 
X & \Ysymm & 1 & \frac{1}{k+1} \\ 
Q & \Yfund & \Yfund & 1-\frac{N+1}{F(k+1)} \\ \hline \hline
& Sp(2\tilde{N}) & SU(2F) & U(1)_R \\ \hline
Y & \Ysymm & 1 & \frac{1}{k+1} \\
q & \Yfund & \overline{\Yfund} & 1-\frac{\tilde{N}+1}{F(k+1)} \\
M_{2n} & 1 & \Yasymm & 2- \frac{2(N+1)-2nF}{F(k+1)} \\
M_{2m+1} & 1 & \Ysymm & 2-\frac{2(N+1)-(2m+1)F}{F(k+1)}
\end{array}\]
Here $\tilde{N}=(2k+1)F-N-2$, and $n=0,\ldots ,k$, $m=0,\ldots ,k-1$.
The superpotential of the magnetic theory is
\beq
\label{wmagn}
W_{magn}= \alpha {\rm Tr}\, Y^{2(k+1)} 
	+\sum_{n=0}^{2k} \beta_{n} M_n q Y^{2k-n}q,
\eeq
where $\alpha$, $\beta_{n}$ are coupling constants.
Now let us perturb this theory by adding a mass term ${\rm Tr}\, X^2$
to the $Sp(2N)$ adjoint of the electric theory. This will break the 
$Sp$ group and make all components of the adjoint massive. With this
superpotential one can have different patterns of symmetry breaking in the 
electric theory, which are non-trivially mapped to the symmetry breaking  
in the magnetic theory.  As described in Ref.~\cite{Spduality}, the 
expectation value for $X$ which has $p_0$ zero eigenvalues 
and $p_l$ eigenvalues $x_l$ breaks the electric $Sp(2N)$ theory to
$Sp(2p_0)\times U(p_1)\times  \ldots U(p_k)$, where $\sum_{j=0}^k p_j=N$.
The magnetic $Sp(2\tilde{N})$ group is broken by the corresponding 
superpotential $\gamma {\rm Tr}\, Y^{2}$ to
$Sp(2(F-p_0-2))\times U(2F-p_1)\times \ldots \times U(2F-p_k)$.
Since we are interested in the case when $Sp(2\tilde{N})\to U(\tilde{N})$,
we choose the values $p_0=F-2$, $p_1=2F-\tilde{N}$ and all other 
$p_i=2F$. This way the magnetic theory after the breaking
becomes an $U(\tilde{N})$ theory, and since the index of embedding 
is two, there are potential $Z_2$ instanton effects in this breaking.
On the electric side with the above values of $p_i$ we find an
$Sp(2F-4)\times U(2F-\tilde{N}) \times U(2F)\times \ldots \times U(2F)$ 
theory, where the $Sp(2F-4)$ has $2F$ fundamentals, and all
unitary factors have $2F$ flavors. Since the  $U(2F)$ factors have 
$2F$ flavors, they confine with a quantum deformed moduli space and 
no superpotential. The magnetic gauge group
$U(\tilde{N})$ is just the dual of the 
$U(2F-\tilde{N})$ factors of the electric side, and the 
superpotential required for duality will be obtained from the 
superpotential of (\ref{wmagn}). However, the electric theory has
an additional $Sp(2F-4)$ gauge group with $2F$ fundamentals, which is
s-confining, that is there is a confining superpotential ${\rm Pf}\, M_0$
generated. What we want to investigate is how this term necessary to 
maintain duality is generated in the magnetic theory. We will see that 
this term is exactly generated by the $Z_2$ instanton.

For this we consider the one instanton effective Lagrangian term of
the original $Sp(2\tilde{N})$ theory. This is given by
\beq 
\label{tHooft1}
\tilde{q}^{2F} \tilde{Y}^{2\tilde{N}+2}\lambda^{2\tilde{N}+2} 
\Lambda^{2\tilde{N}+2-F},
\eeq
where $\tilde{q}$ is the fermionic component of the dual quarks $q$,
$\tilde{Y}$ is the fermionic component of the adjoint $Y$, and $\lambda$ is
the gaugino. The exponents are obtained from counting the zero modes in
the one instanton background. 
Now we show how these zero modes give rise to a superpotential 
coupling in the presence of the expectation value $\langle Y \rangle$.  
Two of the $\tilde{Y}$ are contracted with the mass term $\gamma$.  
The gaugino interaction $Y^{*} \lambda \tilde{Y}$ converts the rest of 
$\tilde{Y}$ together with $2\tilde{N}$ of gaugino fields to ${\langle 
Y\rangle^*}^{2\tilde{N}}$; just like in Section~\ref{SU3}
this is equivalent to the holomorphic part 
$\langle Y\rangle ^{-2\tilde{N}}$ which can appear in the 
superpotential.  Using the superpotential coupling $\beta_{0} M_{0} q 
\langle Y\rangle^{2k} q$, all $\tilde{q}$ but four are contracted as 
$(\beta_{0} M_{0} \langle Y\rangle^{2k})^{2F-2}$. Two out of the remaining
four $\tilde{q}$'s are contracted with the remaining two gauginos 
using the gaugino interaction vertex $q^*  \lambda \tilde{q}$
to $q^*$. These scalars are finally contracted with the remaining two
$\tilde{q}$'s using the superpotential coupling $\beta_{0} M_{0} q 
\langle Y\rangle^{2k} q$ to the fermionic component of 
$(\beta_{0} M_{0} \langle Y\rangle^{2k})$.
Putting everything together, we obtain the superpotential term
\begin{equation}
	\frac{\gamma\mbox{Pf}\,(\beta_{0} M_{0} \langle Y\rangle^{2k})\, 
	\Lambda^{2\tilde{N}+2-F}} {\langle Y\rangle^{2\tilde{N}}} .
\end{equation}
This term is consistent with all symmetries of the theory and has the 
right dimensionality as a superpotential term.  Omitting dependences 
on the mass $\gamma$, the coupling $\beta_0$ 
 and the expectation value $\langle Y \rangle$, 
this is indeed the form expected: $\mbox{Pf}\, M_{0}$.

\section{Conclusions}
We have investigated the question of when instantons in partially broken
gauge groups can have effects on the low-energy effective gauge theory.
We have seen that in some cases (when the embedding of the unbroken 
group $H$ into the original group $G$ is non-trivial) some of the
instantons of the original group $G$ are missing in the low-energy
theory. The effects of these $G/H$ instantons has to be considered
and added to the low-energy effective theory. In the case when
both $G$ and $H$ are simple groups, considering the index of embedding is
sufficient to decide whether such instanton effects may exist or not.
In the more general case one has to consider $\pi_3(G/H)$. 
We have shown several examples of supersymmetric gauge theories 
where these $G/H$ instantons exists and discussed their effects on the
low-energy theory.

\section*{Acknowledgements}

We are grateful to 
Jan de Boer, Bob Cahn, Tohru Eguchi, 
Martin Halpern, Kentaro Hori, Ken Intriligator and Yaron Oz
for useful discussions, and to Ken Intriligator and Witold Skiba
for comments on the manuscript. This work was 
supported in part by the U.S. Department of Energy under Contract 
DE-AC03-76SF00098 and in part by the National Science Foundation under 
grant PHY-95-14797. C.C. is a research fellow of the Miller Institute
for Basic Research in Science. H.M. is an Alfred P. Sloan Foundation
fellow. 

\appendix

\renewcommand{\thesection}{Appendix \Alph{section}}
\section{The Index of Embedding and $\pi_3(G/H)$}
\renewcommand{\theequation}{\Alph{section}.\arabic{equation}}
\setcounter{equation}{0}
\setcounter{footnote}{0}

In this appendix, we show that the index of embedding, defined in the
context of the representation theory, and topology of the coset space
$G/H$ are related.
The statement is the following.
\newtheorem{theorem}{Theorem}
\renewcommand{\thetheorem}{}
\begin{theorem}
Consider a simple compact Lie group $G$ and and its simple subgroup $H$.
Let the index of embedding be $\alpha$.  Then $\pi_3 (G/H) = Z_\alpha$.
\end{theorem}
This is probably a known fact, but we quote our own proof for the sake
of the completeness of the paper.

Here is the proof.  Take a map $S^3 \rightarrow H$ which belongs to
the homotopy class of the generator of $\pi_3 (H)$ ({\it i.e.}\/,
winding number one).  Since $H$ is embedded into $G$, this also
defines a natural map $f: S^3 \rightarrow G$.  The winding number of
the map is computed by
\begin{equation}
  \nu = \frac{1}{24\pi^2} \int_{f(S^3)} \frac{1}{\mu_R (G)} 
  \mbox{Tr}_R (R(g)^{-1} d R(g))^3,
\end{equation}
where the matrix $R(g)$ is in the representation $R$.  The Dynkin index of
the representation $\mu_R (G)$ is needed in this formula to make the
winding number independent of the choice of the representation
$R$.\footnote{The commonly quoted formula (see, {\it e.g.}\/,
  \cite{Jackiw}) does not involve the Dynkin
  index, because it is written with the defining representation of the
  group.}
This can be seen by using the  Maurer--Cartan forms $R(g)^{-1} d R(g) =
\omega^a R(T^a)$ and by rewriting the three-form $\mbox{Tr}_R (R(g)^{-1} d
R(g))^3$ as
\begin{eqnarray}
  \mbox{Tr} (R(g)^{-1} d R(g))^3 
  &=& \mbox{Tr} (R(T^a) R(T^b) R(T^c)) 
  \omega^a \wedge \omega^b \wedge \omega^c \nonumber \\
  &=& \frac{1}{2} \mbox{Tr} (R(T^a) [R(T^b), R(T^c)]) 
  \omega^a \wedge \omega^b \wedge \omega^c \nonumber \\
  &=& \frac{1}{2} f^{bcd} \mbox{Tr} (R(T^a) R(T^d)) 
  \omega^a \wedge \omega^b \wedge \omega^c \nonumber \\
  &=& \mu_R \frac{1}{2} f^{abc} \omega^a \wedge \omega^b \wedge \omega^c .
\end{eqnarray}

Since the map is induced by the embedding of $H$ into
$G$, the group elements $g$ are actually those of $H$:
\begin{equation}
  \nu = \frac{1}{24\pi^2} \int_{f(S^3)} \frac{1}{\mu_R(G)} \mbox{Tr}
  (R(h)^{-1} d R(h))^3. 
\end{equation}
On the other hand, we used the map which winds only once in $H$, and
therefore
\begin{equation}
  1 = \frac{1}{24\pi^2} \int_{f(S^3)} \frac{1}{\mu_R(H)} \mbox{Tr}
  (R(h)^{-1} d R(h))^3,
\end{equation}
where $\mu_R (H) = \sum_i \mu_{R_i} (H)$ is the Dynkin index of the
(in general reducible) representation $R = \sum_i R_i$ of $H$.
Therefore we find $\nu = \sum_i \mu_{R_i}(H)/\mu_R(G) = \alpha$ which
is the index of embedding, and hence the induced map $S^3 \rightarrow
G$ belongs to the homotopy class of $\alpha$ times the generator of
$\pi_3 (G)$. 

The embedding of $H$ into $G$ defines the map $\pi_3 (H) \rightarrow
\pi_3 (G)$ in the exact homotopy sequence
\begin{equation}
  \pi_3 (H) \to \pi_3 (G) \to \pi_3 (G/H) \to \pi_2 (H) =0 ,
\end{equation}
where the generator of $\pi_3(H)$ is mapped to $\alpha$ times the
generator of $\pi_3 (G)$.  Under the assumptions of both $G$ and $H$
being simple, $\pi_3 (H) = \pi_3 (G) = Z$.  Therefore we find
\begin{equation}
  \pi_3 (G/H) = \pi_3(G)/\mbox{Im}(\pi_3(H)) = Z/(\alpha Z) = Z_\alpha.
\end{equation}
This completes the proof.

\section{Explicit Example of a $Z_{2}$ Instanton}
\setcounter{equation}{0}
\setcounter{footnote}{0}

It is probably useful to consider an explicit example of a $G/H$ 
instanton.  Let us take the breaking of $Sp(4)$ to $SU(2)$ with a 
Higgs field in the rank-two symmetric tensor representation.  We 
assume an $N=1$ supersymmetric theory where the potential of the Higgs 
field is the $D$-term potential.  In this case one can write down a 
simple exact solution to Eqs.~(\ref{eq:pureYM2},\ref{eq:Higgs2}).  

To establish the notation, we first write down explicit expression of 
an $SU(2)$ instanton.  
The one-instanton configuration can be constructed as follows.  First take 
the boundary conditions with
\begin{equation}
	U(x) = 
	\frac{t + i \vec{x} \cdot \vec{\sigma}}{(t^{2} + \vec{x}^{2})^{1/2}}, 
	\hspace{1cm}
	A_{\mu} \rightarrow i U \partial_{\mu} U^{\dagger} 
	\mbox{ for } |x|\rightarrow \infty .
\end{equation}
The instanton solution of size $\rho$ is given by
\begin{eqnarray}
	A_{t} &=& \frac{1}{t^{2} + \vec{x}^{2} + \rho^{2}}
		\left( \begin{array}{cc}
			-z & -x+iy\\ -x-iy & z
		\end{array}
		\right), \nonumber \\
	A_{x} &=& \frac{1}{t^{2} + \vec{x}^{2} + \rho^{2}}
		\left( \begin{array}{cc}
			y & t+iz\\ t-iz & -y
		\end{array}
		\right), \nonumber \\
	A_{y} &=& \frac{1}{t^{2} + \vec{x}^{2} + \rho^{2}}
		\left( \begin{array}{cc}
			-x & -it+z\\ it+z & x
		\end{array}
		\right), \nonumber \\
	A_{z} &=& \frac{1}{t^{2} + \vec{x}^{2} + \rho^{2}}
		\left( \begin{array}{cc}
			t & -ix-y\\ ix-y & -t
		\end{array}
		\right) . \label{eq:SU(2)instanton}
\end{eqnarray}
Under this instanton background, the following configuration of the 
Higgs field in the rank-two symmetric tensor representation
\begin{equation}
	H = v \frac{1}{t^{2} + \vec{x}^{2} + \rho^{2}}
	\left( \begin{array}{cc}
	(t + iz)^{2} - (x - iy)^{2} & 2i(tx-yz)\\
	2i(tx-yz) & (t-iz)^{2} - (x+iy)^{2}
	\end{array}
	\right)
\end{equation}
satisfies the boundary condition $H \rightarrow U v U^{T}$, the 
$D$-flatness ({\it i.e.}\/, $V'(H) = 0$), and also $D_{\mu} D_{\mu} H 
= 0$.  Note that the definition of our covariant derivative is 
$D_{\mu} H = \partial_{\mu} H - i A_{\mu} H - i H A_{\mu}^{T}$.  With 
this configuration, the action of the Higgs field is given by 
$8\pi^{2} \rho^{2} v^{2}$.

In the $Sp(4)$ theory, the $Z_{2}$ instanton is given simply by 
embedding the above one-instanton configuration in to an $SU(2)$ 
subgroup, such as
\begin{eqnarray}
	A_{t} &=& \frac{1}{t^{2} + \vec{x}^{2} + \rho^{2}}
		\left( \begin{array}{cccc}
			-z & 0 & -x+iy & 0\\ 0 & 0 & 0 & 0\\
			-x-iy & 0 & z & 0\\ 0 & 0 & 0 & 0
		\end{array}
		\right), \; \; \mbox{etc.} \nonumber \\
	H &=& v 
	\left( \begin{array}{cccc}
	\frac{(t + iz)^{2} - (x - iy)^{2}}{t^{2} + \vec{x}^{2} + \rho^{2}} & 0 
	& \frac{2i(tx-yz)}{t^{2} + \vec{x}^{2} + \rho^{2}} & 0\\
	0 & 1 & 0 & 0\\
	\frac{2i(tx-yz)}{t^{2} + \vec{x}^{2} + \rho^{2}} & 0 & 
	\frac{(t-iz)^{2} - (x+iy)^{2}}{t^{2} + \vec{x}^{2} + \rho^{2}} & 0\\
	0 & 0 & 0 & 1
	\end{array}
	\right).
\end{eqnarray}
This Higgs field configuration is $D$-flat and satisfies $D_{\mu} 
D_{\mu} H = 0$, {\it i.e.}\/, a solution to the equation of motion.
Two-instantons, however, belong to a topologically trivial class in 
$\pi_{3} (Sp(4)/SU(2))$.  For instance, take the following two-instanton 
configuration of the gauge field\footnote{Note that 
if $A_{\mu}^{SU(2)}$ is an instanton, $-(A_{\mu}^{SU(2)})^{T}$ is also 
an instanton rather than an anti-instanton.  This can be checked 
easily by calculating all field strengths and verify the self-duality.}
\begin{equation}
	A_{\mu} = 
		\left( \begin{array}{cc}
			A_{\mu}^{SU(2)} & 0 \\ 0 & -(A_{\mu}^{SU(2)})^{T}
		\end{array}
		\right),
\end{equation}
where $A_{\mu}^{SU(2)}$ are the two-by-two matrices given in 
Eq.~(\ref{eq:SU(2)instanton}).  Then 
a trivial Higgs field configuration
\begin{equation}
	H = v \left( \begin{array}{cc}
			0 & 1 \\ 1 & 0
		\end{array}
		\right) ,
\end{equation}
where $1$ is a two-by-two unit matrix, satisfies $D_{\mu} H = 0$ and the 
boundary condition, as seen in
\begin{equation}
	H = v \left( \begin{array}{cc}
		U & 0\\0 & U^{*}
	\end{array}
	\right)
	\left( \begin{array}{cc}
			0 & 1 \\ 1 & 0
		\end{array}
		\right)
	\left( \begin{array}{cc}
		U^{T} & 0\\0 & U^{\dagger}
	\end{array}
	\right)
	= v \left( \begin{array}{cc}
			0 & 1 \\ 1 & 0
		\end{array}
		\right).
\end{equation}
Therefore this two-instanton configuration is nothing but the 
one-instanton of the low-energy $SU(2)$ theory.

\end{document}